\documentclass{ws-procs961x669}            

\newcommand{\la}{\langle}
\newcommand{\ra}{\rangle}
\newcommand{\be}{\begin{equation}}
\newcommand{\ee}{\end{equation}}
\newcommand{\bea}{\begin{eqnarray}}
\newcommand{\eea}{\end{eqnarray}}
\newcommand{\bes}{\begin{subequations}}
\newcommand{\ees}{\end{subequations}}
\newcommand{\w}{\omega}
\newcommand{\kb}{\kappa_b}
\newcommand{\kc}{\kappa_c}

\def\fbh{ {\cal H}_b^+ }

\def\fch{ {\cal H}_c^+ }

\def\kn{\kappa_N}

\begin{document}
\title{Linear growth of the two-point function for the Unruh state in $1+1$ dimensional black holes}

\author{Paul R. Anderson$^*$ and Zachary P. Scofield}

\address{Department of Physics, Wake Forest University,\\
Winston-Salem, NC 27109, USA \\
$^*$E-mail: anderson@wfu.edu}

\author{Jennie Traschen}

\address{Department of Physics, University of Massachusetts,\\
Amherst, MA  01003, USA\\
E-mail: traschen@umass.edu}

\begin{abstract}
The symmetric two-point function for a massless, minimally coupled scalar field
in the Unruh state is examined for Schwarzschild-de Sitter spacetime in two dimensions.
This function grows linearly in terms of a time coordinate that is well-defined on the future black hole and cosmological horizons, when the points are split in the space direction.  This type of behavior also occurs in two dimensions for other static black hole spacetimes
when the field is in the Unruh state, and at late times it occurs in spacetimes where a black hole forms from the
collapse of a null shell. The generalization to the
case of the symmetric two-point function in two dimensions for a massive scalar field in Schwarzschild-de Sitter spacetime is discussed.
\end{abstract}


\bodymatter

\section{Introduction}\label{aba:sec1}

Hawking~\cite{Hawking:1974sw} predicted that black holes that form from collapse evaporate and at late times the particles are in a thermal distribution with a temperature that is  proportional to the surface gravity of the black hole.  In four dimensions (4D) it is very difficult to compute quantities such as the two-point function and the stress-energy tensor for a quantum field in the spacetime of a black hole that forms from collapse. Instead, it is significantly easier, although not easy, to do such computations in eternal black hole spacetimes which are either static or stationary.  It is easier yet to work with a massless minimally coupled scalar field in spacetimes with two dimensions (2D) that have similar structures to the 4D black hole spacetimes.

Unruh~\cite{Unruh:1976db} showed that, for an isolated eternal black hole in an asymptotically flat spacetime, there is a particular state for a quantum field that has the same properties as the {\it in} state for that field has at late times in a spacetime where the black hole forms from collapse. In particular, this state, called the Unruh state,  has the same flux of particles at infinity for an eternal black hole as occurs at late times for a black hole that forms from collapse.  Thus computations of various quantities for a quantum field in the Unruh state can give insight into the late time behaviors of these quantities in spacetimes where a black hole forms from collapse.

The simplest 4D black hole spacetimes to compute quantum effects in are those which are static and spherically symmetric outside of the event horizon where the metric has the general form
\be ds^2 = -f(r) dt^2 + f^{-1}(r) dr^2 + r^2 d \Omega^2 \;. \label{metric-4Dssss} \ee
These are easily changed to 2D spacetimes by dropping the last term on the right.  All 2D spacetimes are conformally flat and the massless, minimally coupled scalar field is conformally invariant in such spacetimes making it possible to analytically solve the mode equation in coordinates for which
the metric is conformally flat.  

In Ref.~\citenum{and-tra}, the Hadamard Green's function, which is the symmetric two-point function, was computed and studied for a massless, minimally coupled scalar field in the Unruh state in 2D for Schwarzschild-de Sitter(SdS) spacetime, Schwarzschild spacetime, de Sitter space, a class of Bose-Einstein condensate analog black hole spacetimes, and a spacetime in which a null shell collapses to form a black hole.  It was shown that in all of these cases there is linear growth in terms of a time coordinate $T$ that is regular on the future horizon(s) when the points are split in the spatial direction.

In this proceeding we summarize some of the results of Ref.~\citenum{and-tra} for black holes.  We also discuss the computation of the symmetric two-point function for a massive, minimally coupled scalar field in SdS.  In this case, the mode functions and the two-point function must be computed numerically.

In Sec. 2, some properties of 2D SdS are reviewed along with the Unruh state for a massless, minimally coupled scalar field.
The Hadamard Green's function for this field is displayed in Sec.~3 and its linear growth in time is discussed.
Ongoing work related to the massive scalar field in 2D SdS is discussed in Sec.~4.  In Sec.~5 the computation of the symmetric two-point function
for a massless, minimally coupled scalar field in a 2D collapsing null shell spacetime is reviewed.  Our results are summarized in Sec.~6.
Throughout, our units are $\hbar = c = G = 1$.

\section{Massless, minimally coupled scalar field in SdS}

\begin{figure}[h]
\begin{center}

\includegraphics[trim=1cm 17cm 1cm 1cm, clip=true, totalheight=0.25\textheight]{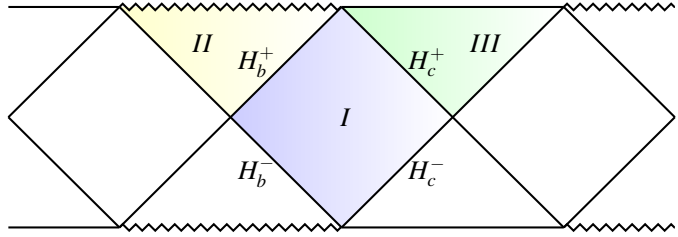}
\end{center}
\caption{Penrose diagram for SdS.  Region I is the static patch between the black hole horizons and cosmological horizons, Region II is the interior of the black hole, and Region III is the cosmological region.}
\label{fig:SdS}
\end{figure}
A Penrose diagram for SdS is given in Fig.~\ref{fig:SdS}.
The 2D SdS metric is given by
\bes \bea ds^2 &=& -f(r) dt^2 + \frac{dr^2}{f(r)}\;, \label{metric-SdS-2D} \\
f(r)&=&1-\frac{2M}{r}-  H^2 r^2   = \ - \frac{H^2}{r} (r-r_c ) (r-r_b )(r +r_c +r_b ) \;. \label{f-Sds}
\eea \ees
Here $r_c > r_b $ are the locations of the cosmological and black hole horizons respectively, and
 $ \frac{1}{3}\Lambda =  H^2$. The two parameterizations are related by
 \be
 M = {r_br_c(r_b+r_c)\over 2 (r_b^2+r_c^2+r_br_c)},\qquad H^2 = {1\over r_b^2+r_c^2+r_b r_c}  \label{M-H2}
\ee
It is useful to define the tortoise coordinate
\be\label{rstar1}
 r_* (r)  =  \int {dr\over f} \;.
\ee
The ingoing and outgoing radial null coordinates are
\be\label{uvdef}
u= t-r_* \ , \quad v=t+r_* \;.
\ee
One can define two sets of Kruskal coordinates, one for each horizon. The ones we are concerned with are
\bea\label{UVuv}
 U_b  &= &  \frac{1}{\kappa_b} e^{-\kappa_b u} \   , \ \   r<r_b \ , \quad
 U_b  = - \frac{1}{\kappa_b} e^{-\kappa_b u} \ , \ \  r>r_b \\  \nonumber
V_c & = &- \frac{1}{\kappa_c} e^{-\kappa_c v}  \  ,\ \  r<r_c \ , \quad
 \quad V_c =  \frac{1}{\kappa_c} e^{-\kappa_c v}  \ \  , \ \   r>r_c  \;.
  \eea
The surface gravities for the two horizons are
  \bea\label{surfgravs}
  \kb &=& {H^2 \over 2 r_b } (r_c -r_b ) (r_c + 2r_b )  \;, \nonumber  \\
    \kc &=& {H^2 \over 2 r_c } (r_c -r_b ) (2r_c + r_b )  \;,
      \eea
where for a horizon at $r = r_h$ we set $2\kappa _h = |f' (r_h )| $ so that each of the surface gravities is a positive quantity.

A set of coordinates that are good across $\fbh$ and $\fch$
 was found in Refs.~\citenum{Gregory:2017sor,Gregory:2018ghc}. Let
 \be\label{coord1}
 T = t + h(r ) \ ,\ \quad {\rm where} \quad { dh \over dr}  = {j \over f} \ , \quad  \quad j (r) =-\gamma r +{\beta \over r^2 } \;,
 \ee
and
 \be\label{gammabeta}
\gamma = {r_c^2 +r_b^2 \over r_c^3 - r_b^3 } \ , \quad \beta = {r_c^2 r_b^2 (r_b + r_c ) \over r_c^3 - r_b^3 } \;.
\ee
Then the 2D SdS metric becomes
\be\label{regularmetric}
ds^2   =  -f(r)  dT^2 + 2j(r) dr dT +{1-j^2 \over f} dr^2 \;.
\ee
Note that for these choices of the constants $\gamma$ and $\beta$, one finds that $j (r_b ) = 1$ and $j (r_c ) =-1$, which
ensures that $T$ interpolates between the ingoing null coordinate $v$  at
the future black hole horizon and the outgoing null coordinate $u$ at the future cosmological horizon. These are
ingoing and outgoing Eddington-Finklestein coordinates respectively.
The metric (\ref{regularmetric})
is well behaved in the static and cosmological regions and has the Eddington-Finklestein form on both the future black hole and future cosmological horizons. The coordinate $T$ stays timelike and $r$ stays spacelike beyond the cosmological horizon.

The definitions of $r_*$ and $T$  each contain an arbitrary constant which can be chosen such that
\be\label{Tbc}
 T=u \  {\rm on} \ \fch \  ,\quad \quad {\rm and} \quad  T=v  \ {\rm on} \ \fbh  \;.
 \ee
 The result is\cite{and-tra}
\bea\label{Tcoord}
T=  t  &+ & h(r)  \\ \nonumber
= t  &+ &
{1\over 2\kb} \log  {| r-r_b | \over r_c - r_b  } +{1\over 2\kc} \log  {| r-r_c | \over r_c -r_b  } +
{1\over 2}\left( {r_c \over  r_b \kb }   -{1\over \kn}\right) \log   { r +  r_c + r_b \over  r_c + 2r_b}  \\ \nonumber
 &- & {r_b r_c\over 2(  r_c - r_b)} \log {r^2 \over r_c r_b}  +{r_c \over 4 r_b \kb } \log { r_c + 2 r_b\over 2 r_c +  r_b }  \;,
\eea
and
\bea\label{rstar}
 r_* (r) &  =& {1\over 2\kb} \log  {| r-r_b | \over r_c - r_b }- {1\over 2\kc} \log  {| r-r_c | \over r_c  -r_b}
 + {1\over 2\kn} \log   {| r+r_c + r_b  | \over r_c + 2r_b } \\ \nonumber
 & - & {r_c \over 4r_b \kb } \log{2r_c + r_b \over r_c +2r_b}  -{r_b r_c \over 2 ( r_c - r_b )}\log{r_b\over r_c} \;,
\eea
where
 \be \kn = {H^2 \over 2(r_c +  r_b ) } (2r_c + r_b ) (r_c + 2r_b ) \;. \ee

 To relate the Kruskal coordinates to $T$ and $r$, note that
 $u= T-( h(r) +r_* ) $ and $v= T- h(r) + r_*  $. Substitution gives
 \be\label{vutr}
  V_{c} = {1\over \kc}  e^{-\kc T}  \tilde{V}_c   \ ,\  \quad {\rm and} \quad
 U_{b} ={1\over \kb} e^{-\kb T}  \tilde{U}_b \;,
\ee
where $\tilde{V}_c$ and $\tilde{U}_b$ only depend on $r$ and are given by
\bea\label{krutor}
 \tilde{V}_c &=&  e^{\kc (  h -r_* ) } =  { r-r_c  \over r_c -r_b}
\left({ r+r_c +r_b \over  r_c + 2 r_b}  \right)^{r_b / 2r_c } \left( { r_b\over r} \right)^{H^2 r_b (2 r_c +r_b )/2 }  \;,
\\ \nonumber
 \tilde{U}_b &=& - e^{\kb ( h +r_* ) } = -  {  ( r-r_b ) \over  r_c - r_b}
\left({ r+r_b +r_c \over  2r_c + r_b}  \right)^{r_c / 2r_b } \left( { r_c\over r} \right)^{ H^2 r_c ( r_c +2 r_b ) /2} \;.
\eea
Note that these expressions work both inside and outside the black hole and cosmological horizons.

The complete set of modes that comprise the Unruh state consist of modes that are positive frequency with respect to $U_b$ on
the past black hole horizon and modes that are positive frequency with respect to $V_c$ on the past cosmological horizon~\cite{Tadaki:1990a,Tadaki:1990b,Markovic:1991ua}.
\be  p^b_\w = \frac{e^{-i \w U_b(u)}}{\sqrt{4 \pi \w}}
 \;\;{\rm and}\;\; p^c_\w = \frac{e^{-i \w V_c(v)}}{\sqrt{4 \pi \w}} \label{Unruh-modes} \ee
Expanding the massless minimally coupled scalar field in terms of these gives
\be \phi = \int_0^\infty d \w \left[ a^b_\w p^b_\w + a^c_\w p^c_\w +
 a^{b\dagger}_\w p^{b*}_\w + a^{c\dagger}_\w p^{c*}_\w \right] \;, \ee
where $a^{(b,c)}$ are the annihilation operators.

\section{Symmetric Two-Point Function in SdS}

For a massless, minimally coupled scalar field in 2D SdS in the Unruh state, the symmetric two-point function for an arbitrary splitting of the points is\cite{and-tra}
\begin{align}
  G^{(1)}(x,x') &=  \la U | \left(\phi(x) \phi(x')+\phi(x')\phi(x)\right)| U \ra \nonumber \\
 &= \int_0^\infty d\w [ p^b_\w(x) p^{b \, *}_\w(x') + p^c_\w(x) p^{c \, *}_\w(x') + {\rm c.c.} ] \nonumber \\
   &= \frac{1}{2 \pi} \int_{\omega_0}^\infty \frac{d \w}{\w} \left\{ \cos[\w (U_b - U_b^{'})] + \cos[\w(V_c - V_c^{'})] \right\} \nonumber \\
    &= - \frac{1}{2 \pi} \left\{ ci[\omega_0 |U_b - U_b^{'}|] + ci[\omega_0 |V_c - V_c^{'}|] \right\} \;.
    \end{align}
Here $ci$ is the cosine integral function which has the expansion $ci(z) = \gamma_E + \log z + O(z^2)$,  and $\omega_0$ is a small infrared cutoff.

Let $x$ be on $H^b_+$ and $x'$ on $H^c_+$ and take $T'=T$.  Note that
 on $H^{+}_b$, $U_b = 0$, $v = T $ and on $H^+_c$, $V_c = 0$, $u = T $.  Then
 \begin{align}
 U_b - U_b^{'} & =    \ - \kb^{-1} e^{-\kb T} \;, \qquad
 V_c - V_c^{'}  =     \ \kc^{-1} e^{-\kc T } \;,  \nonumber \\
2\pi G^{(1)}(T,r;T ,r' ) &=  (\kb + \kc) T  - \log \left[  (\kb \kc)^{-1} \omega_0^2  \right]  -2\gamma_E  \;. \end{align}
Thus one sees an unexpected linear growth in the time coordinate $T$ for this separation of the points.
 For a more general separation
\begin{align} 2\pi G^{(1)}(T,r ; T, r' ) &= T (\kappa_b +\kappa_c ) -  \log\left(  {\omega_0^2 \over \kappa_b\kappa_c } |  \Delta \tilde{U}_b  \Delta \tilde{V}_c |   \right) -2\gamma_E  \;, \end{align}
where $ \Delta \tilde{U}_b $ and $\Delta \tilde{V}_c$ are functions of $r$ and $r'$, but not $T$.

In Ref.~\citenum{and-tra} a similar linear growth in a time coordinate for $G^{(1)}(x,x')$ in 2D for a massless minimally coupled scalar field was found in Schwarzschild spacetime for the Unruh state, de Sitter spacetime for the Bunch-Davies state, and a class of Bose-Einstein condensate analog black hole spacetimes for the Unruh state.
A general argument was also given that linear growth should occur whenever there is a past horizon and no scattering for the Unruh state.

One may ask what these situations have in common.  In each case the symmetric two-point function is computed for a massless minimally
coupled scalar field in 2D in the Unruh state or its analog, the two-point function has an infrared divergence which requires an infrared cutoff, and each spacetime has a static patch and at least one past horizon.  In the next two sections, two cases are discussed in which one or more of these conditions does not occur.

\section{Massive Scalar Field in 2D SdS}

One way to investigate the generality of the linear growth in time is to work with a massive scalar field.  Work is in progress to compute the symmetric two-point function for this field.  It is significantly more difficult than for the massless scalar field because the mode equation
contains an effective potential term and does not have simple closed form solutions.  Instead, when separation of variables is used, the radial
part of the mode equation can be solved numerically.

The mode equation cannot be separated in Kruskal coordinates but it can be separated in terms of the coordinates $t$ and $r$ in~\eqref{f-Sds}. The mode equation in this case is
\be - \frac{\partial^2 h}{\partial t^2} + \frac{\partial^2 h}{\partial r_*^2} - m^2 f  = 0 \;.  \ee
The relevant solutions are of the form
\be h(t,r) = \frac{e^{-i \w t}}{\sqrt{4 \pi \w}}  \chi_\w(r)  \;, \ee
with
\be   \frac{d^2 \chi}{d r_*^2} +(\w^2 - m^2 f) h = 0 \;. \ee
A complete set of modes that specify the Boulware state can be obtained by combining modes which on the past black hole horizon have the form
\be h^b_\w = \frac{e^{-i \w u}}{\sqrt{4 \pi \w}} \;, \ee
and those which on the past cosmological horizon have the form
\be h^c_\w = \frac{e^{-i \w v}}{\sqrt{4 \pi \w}} \;. \ee
The modes that specify the Unruh state can be expanded in terms of these modes with the result
\be p^{(b,c)}_\w = \int_0^\infty d \w' \,[\alpha^{(b,c)}_{\w \w'} h^{(b,c)}_{\w'} + \beta^{(b,c)}_{\w \w'} h^{{(b,c)}\, *}_{w'} ]  \;. \label{pbcw} \ee
Since the mode equation reduces to its massless form on the past horizons, the initial conditions for the Unruh state are exactly the same as
in the massless case~(\ref{Unruh-modes}).
Using the usual scalar product~\cite{b-d-book}, one finds~\cite{and-gho}
\bea \alpha^{(b,c)}_{\w \w' } &=&  \frac{1}{2 \pi} \sqrt{\frac{\w'}{\w}} \, (\kappa_{(b,c)})^{-1 -i \frac{ \w'}{\kappa_{(b,c)}}} \frac{\Gamma( - i \frac{\w'}{\kappa_{(b,c)}})}{ (-i \w)^{-i \frac{ \w'}{\kappa_{(b,c)}}}} \;, \nonumber \\
\beta^{(b,c)}_{\w \w'} &=&  \frac{1}{2 \pi} \sqrt{\frac{\w'}{\w}} \, (\kappa_{(b,c)})^{-1 + i \frac{\w'}{\kappa_{(b,c)}}} \frac{\Gamma( i \frac{\w' }{\kappa_{(b,c)}})}{ (-i \w)^{i \frac{\w'}{\kappa_{(b,c)}}}} \;. \label{alpha-beta-bc}
\eea
Note that these are divergent in the limit $\w'\to 0$ due to the poles in the Gamma functions.  However, we have shown that because of scattering
effects, these divergences are canceled in the integrand of Eq.~(\ref{pbcw}).
Also, it is important to note that the limit $\w' \to 0$ needs to be taken before the limits $r \to r_b$ and $r \to r_c$.  They cannot be interchanged.

At present, work is in progress to compute $p^b_\w$ and $p^c_\w$ numerically.  It is not yet known whether the divergence that exists for the massless case for these mode functions continues to be present in the massive case.

\section{Collapsing null shell spacetime}

\begin{figure}[h]
\begin{center}
\includegraphics[trim=1cm 20cm 0cm 1cm, clip=true, totalheight=0.25\textheight]{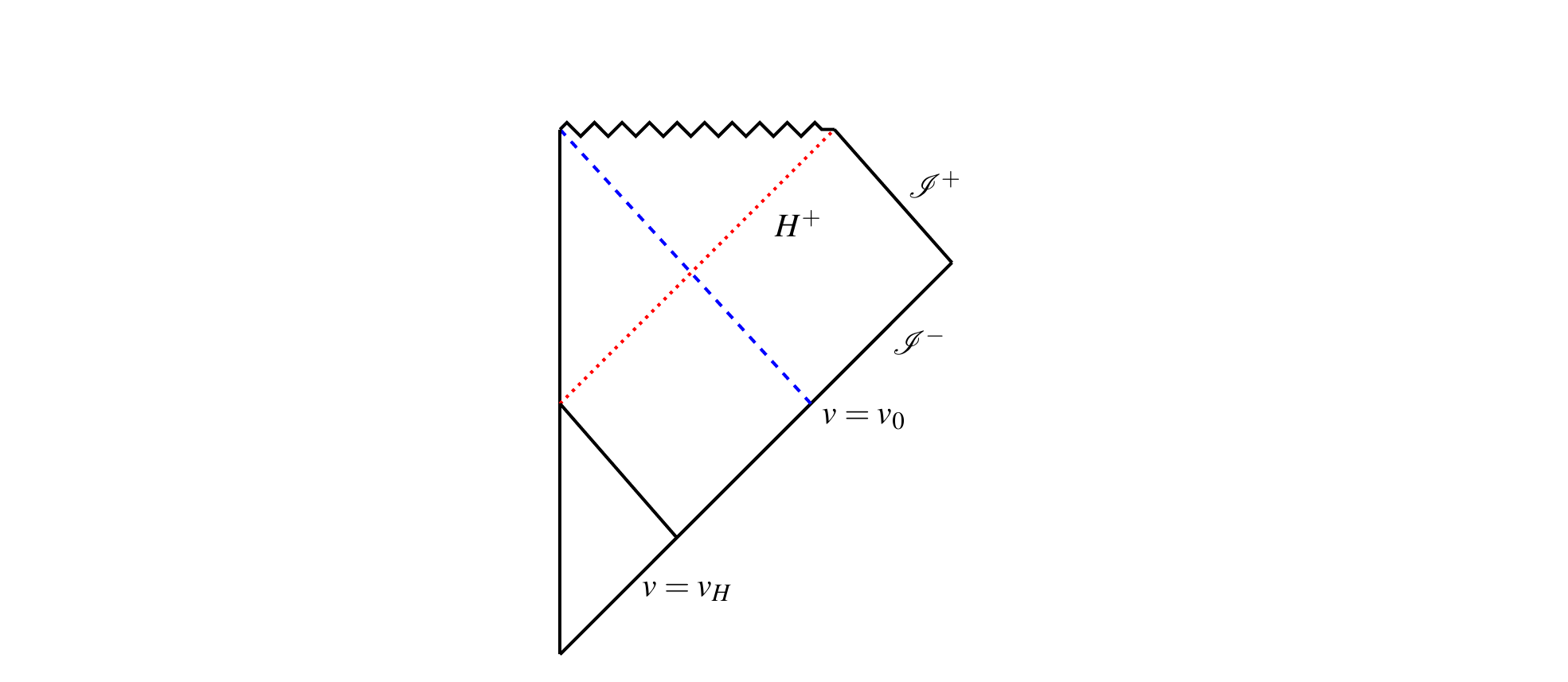}
\end{center}
\caption{Penrose diagram for a 2D spacetime in which a null shell collapses to form a black hole and there is a perfectly reflecting mirror at the origin.  The trajectory of the shell is given by the dashed blue line and the event horizon is given by the dotted red line.}
\label{fig:null-shell}
\end{figure}

Another calculation that was done in Ref.~\citenum{and-tra} was to compute the two-point function for
 a massless scalar
field in a 2D spacetime in which a null shell collapses to form a Schwarzschild black hole.  Inside the shell the metric
is the flat space metric
\be ds^2 = -dt^2 + dr^2 \;, \ee
and outside it is Eq.~(\ref{f-Sds}) with $H = 0$, which is the Schwarzschild metric.
If one puts in a perfectly reflecting mirror and requires that the field vanish at the surface of the mirror then
one obtains the same conditions on the mode functions at $r=0$ as in the 4D case.  The Penrose diagram for this spacetime is shown in Fig.~\ref{fig:null-shell}.

Inside the shell, the usual null coordinates are
\be u = t -r \;, \qquad v = t + r \;, \ee
and outside the shell they are
\be u_s = t_s -r_* \;, \qquad v = t_s + t_* \;. \ee
The coordinates $r$ and $v$ are continuous across the trajectory of the shell which is $v = v_0$, for some $v_0$.
At the null shell surface one finds~\cite{m-p,Fabbri:2005mw}
\be  u(u_s) = v_H - 4 M W\left[\exp\left(\frac{v_H-u_s}{4 M}\right)\right] \;, \label{u-us} \ee
where $v_H = v_0 - 4 M$ and $W$ is the Lambert W function.  On past null infinity both $u_s$ and $u$ are equal to $-\infty$.  On the future horizon of the black hole $u_s = \infty$ and $u = v_H$.

Because of the mirror, the mode functions must vanish at $r = 0$.  For the {\it in} state, at past null infinity they have the form
\be f^{\rm in}_\w \to \frac{e^{-i \w v}}{\sqrt{4 \pi \w}} \;. \label{fin-null}\ee
The solution that satisfies these two conditions is
\be f^{\rm in}_\w = \frac{1}{\sqrt{4 \pi \w}} (e^{-i \w v} - e^{-i \w u}) \;. \ee
Since there is no scattering in 2D for the massless, minimally coupled scalar field, outside the shell the solution is
\be f^{\rm in}_\w = \frac{1}{\sqrt{4 \pi \w}} (e^{-i \w v} - e^{-i \w u(u_s)}) \;. \ee

 Note that $f^{\rm in}_\w = 0$ at $\w = 0$.  Thus there is no infrared divergence in the two-point function for this state and no infrared cutoff
 is necessary.  One finds that\cite{and-tra}
 \be 2 \pi  G^{(1)}(x,x')= \log \left[\frac{|v-u(u_s')||v'-u(u_s)|}{|v-v'||u(u_s)-u(u_s')|} \right] \;. \ee
Using $v = T - h(r) + r_*$ and $u = T - h(r)-r_*$, one finds for $T = T'$ and fixed values of $r$ and $r'$, that
at late times
\be  2 \pi G(T,r;,T,r') \to \kappa T + \log (\kappa^2 T^2)  + (r, r' {\rm dependent \; terms}) \;. \ee
Thus, there is not only a linear growth in $T$ as there is for the Unruh state, but there is also a secondary logarithmic growth.

\section{Conclusions}

 In Ref.~\citenum{and-tra}, the behavior of the two point function was investigated for a massless, minimally coupled scalar field in the Unruh state in certain 2D spacetimes.  It was found that linear growth in time of the form
 \be 2 \pi G^{(1)}(T,r;T,r') = T \sum_J \kappa_J  + (r, r' {\rm dependent \; terms}) \label{G-T-gen} \ee
 occurs, where the sum is over the surface gravities of the past horizons and $T$ is a time coordinate that is regular on the future horizons.  Along with the existence of one or two past horizons, the Green's function for each spacetime is infrared divergent and thus requires an infrared cutoff to regularize this divergence.

 The two-point function for this field in a 2D spacetime  in which a null shell collapses to form a black hole was also computed.
In this case there is no past horizon and thus no Unruh state.  Instead, since the spacetime is asymptotically flat, the state for the massless, minimally coupled scalar field was chosen to be the initial vacuum state.
By putting a perfectly reflecting mirror at the origin and requiring that the mode functions vanish there, the same type of condition that one finds in the 4D collapsing null shell spacetime if the shell is spherically symmetric was imposed.  As a result of this condition, the two-point function has no infrared divergences.

It was found that at early times there is no linear growth in time of the form~(\ref{G-T-gen}), however at late times there is growth of this form.  There is also a subleading term that diverges logarithmically in time at late times.  Thus, the linear growth in time found in the eternal black hole calculations also occurs, at least in 2D, for a black hole that forms from the collapse of a null shell.  Perhaps this is not surprising since the Unruh state is supposed to give the leading order, late time behavior of quantities such as the two-point function and stress-energy tensor for the quantum field in the case of a black hole that forms from collapse.  However, it does show that the linear growth in time does not depend on
the existence of either a past horizon or an infrared divergence in the two-point function.

To see whether the linear growth in time only occurs for the massless scalar field, the case of a massive scalar field in the Unruh state in 2D SdS is being investigated.  It has been shown that, when the mode functions for the Unruh state are expanded in terms of those for the Boulware state, an infrared diverge that occurs in the massless case is not present in the massive one.  Work is in progress to determine whether the linear growth in time of the two-point function also occurs for a massive scalar field.

\section{Acknowledgments}
P.R.A. and J.T.  would like to thank David Kastor for useful conversations. P.R.A. would also like to thank Alessandro Fabbri and Ian Moss for useful conversations.  This work came about as a result of conversations at the NORDITA
Gravitational Physics with Lambda Workshop 2018.
 We would like to thank NORDITA for their generous support.  This work was supported in part by the National Science Foundation under Grants No. PHY-1505875 and PHY-1912584 to Wake Forest University.

\end{document}